\documentclass{SolarPhysics}    % Specifies the document style.
\usepackage[optionalrh]{spr-sola-addons}
\usepackage{epsfig}
\usepackage{url}
\usepackage{fixltx2e}
\usepackage[usenames]{color}

\newcommand{\gapprox}{\lower.4ex\hbox{$\;\buildrel >\over{\scriptstyle\sim}\;$}}
\newcommand{\lapprox}{\lower.4ex\hbox{$\;\buildrel <\over{\scriptstyle\sim}\;$}}

\def\etal{{\it et al.}}
\def\eg{{\it e.g.,~}}
\def\ie{{\it i.e.,~}}

\def\aap  {{\sl Astron. Astrophys.}\ }   % Astronomy and Astrophysics
\def\apj  {{\sl Astrophys. J.}\ }        % The Astrophysical Journal
\def\apjl {{\sl Astrophys. J. Lett.}\ }  % The Astrophysical Journal Lett.
\def\aj   {{\sl Astronom. J.}\ } 	 % The Astronomical Journal
\def\sp   {{\sl Solar Phys.}\ }          % Solar Physics
\begin{document}
\begin{article}
\begin{opening}
\title{A Nonlinear Force-Free Magnetic Field Approximation 
	Suitable for Fast Forward-Fitting to Coronal Loops. 
	II. Numeric Code and Tests}

\author{Markus J. Aschwanden and Anna Malanushenko}
\runningauthor{M.J. Aschwanden and A. Malanushenko}
\runningtitle{Nonlinear Force-Free Magnetic Field}

\institute{Solar and Astrophysics Laboratory,
	Lockheed Martin Advanced Technology Center, 
        Dept. ADBS, Bldg.252, 3251 Hanover St., Palo Alto, CA 94304, USA; 
        (e-mail: \url{aschwanden@lmsal.com})}

\date{Received 30 Nov 2011; Revised 3 July 2012; Accepted ...}

\begin{abstract}
Based on a second-order approximation of nonlinear force-free 
magnetic field solutions in terms of uniformly twisted field lines
derived in Paper I, we develop here a numeric code that is capable
to forward-fit such analytical solutions to arbitrary magnetogram
(or vector magnetograph) data combined with (stereoscopically
triangulated) coronal loop 3D coordinates. We test the code here
by forward-fitting to six potential field and six nonpotential field
cases simulated with our analytical model, as well as by forward-fitting
to an exactly force-free solution of the Low and Lou (1990) model.
The forward-fitting tests demonstrate: (i) a satisfactory convergence 
behavior (with typical misalignment angles of $\mu \approx 1^\circ-10^\circ$),
(ii) relatively fast computation times (from seconds to a few minutes),
and (iii) the high fidelity of retrieved force-free $\alpha$-parameters
($\alpha_{\rm fit}/\alpha_{\rm model} \approx 0.9-1.0$ for simulations
and $\alpha_{\rm fit}/\alpha_{\rm model} \approx 0.7\pm0.3$ for the Low and Lou
model). The salient feature of this numeric code is the relatively
fast computation of a quasi-forcefree magnetic field, which closely
matches the geometry of coronal loops in active regions, and 
complements the existing {\sl nonlinear force-free field (NLFFF)} 
codes based on photospheric magnetograms without coronal constraints.
\end{abstract}

\keywords{Sun: Corona --- Sun: Magnetic Fields}

\end{opening}

\section{	Introduction 				}

This paper contains a description of a new numerical code that performs 
fast forward-fitting of {\sl nonlinear force-free magnetic fields}
(NLFFF). An alternative NLFFF forward-fitting code has been pioneered by 
Malanushenko \etal (2009), which first fits separate linear force-free 
solutions to individual loops, and in a next step retrieves a self-consistent
NLFFF solution from the obtained linear force-free $\alpha$-values 
(Malanushenko \etal, 2011). 
Since any calculation of a single NLFFF solution requires substantial 
computing time, we expore here a much faster NLFFF forward-fitting code 
that retrieves a self-consistent quasi-forcefree magnetic field with 
somewhat reduced accuracy (\ie second order in $\alpha$), but should 
be still sufficient for most practical applications.  

The NLFFF models are thought to describe the magnetic field in the
solar corona in a most realistic way, because the required force-freeness
and divergence -freeness fulfill Maxwell's electrodynamic equations for a 
steady-state situation. Except for very dynamic episodes, such as flares
or magnetic reconnection events, the magnetic field corona is thought to 
evolve close to a force-free steady state. NLFFF models reveal also the
magnitude and topology of field-aligned currents, which are crucial for
undestanding energetic processes in the solar corona.

About a dozen NLFFF codes exist that have been described in
detail and quantitatively compared (Schrijver \etal, 2005, 2006; Metcalf
\etal, 2008; DeRo\-sa \etal, 2009), which includes: (i) divergence-free and
force-free optimization algorithms (Wheatland \etal, 2000;
Wiegelmann, 2004), (ii) the evolutionary magneto-frictional method
(Yang \etal, 1986; Valori \etal, 2007), or a Grad-Rubin-style (Grad and Rubin,
1958) current-field iteration method (Amari \etal, 2006; Wheatland, 2006;
Malanushenko \etal, 2009).
Most of these NLFFF algorithms are using a photospheric boundary condition
(in form of a magnetogram or 3D vector magnetograph data) and extrapolate
the magnetic field in a coronal box above the photospheric boundary, by
optimizing the conditions of divergence-freeness and force-freeness
(for a general overview of non-potential field calculation methods see, \eg 
Aschwanden, 2004). Only the code of Malanushenko \etal (2009)
uses loop coordinates as additional constraints from the coronal volume.
The methods have different degrees of accuracy, which can be quantified
by an average misalignment angle between the theoretical model and
observed (stereoscopically triangulated) coronal loops, which typically
amounts to $\mu \approx 24^\circ-44^\circ$ (see Table 1 in 
DeRosa \etal, 2009). These NLFFF codes are relatively computing-intensive
(with typical computation times of several hours to a over a day), and
thus are not suitable for forward-fitting, which requires many iterations.

In Paper I (Aschwanden, 2012) we derived an approximation of a general 
solution of a class of NLFFF models (with twisted magnetic fields) 
that is suitable for fast forward-fitting
to coronal loops. The accuracy of this ``quasi-NLFFF solution'' is of
second-order in the force-free parameter $\alpha$. Obviously, we have
a trade-off between accuracy and computation speed. This fast forward-fitting
code can be applied to virtually every kind of simulated or observed
magnetogram or 3D vector magnetograph data, combined with constraints from 
coronal loop coordinates,
in form of 2D or 3D coordinates as they can be obtained by stereoscopic
triangulation (\eg Feng \etal, 2007a; Aschwanden \etal, 2008a). 
In this Paper II we describe this first ``fast'' NLFFF forward-fitting code
and test it with simulated data and analytical NLFFF solutions, such as
obtained from the Low and Lou (1990) model.

\clearpage

\begin{figure}
\centerline{\includegraphics[width=1.0\textwidth]{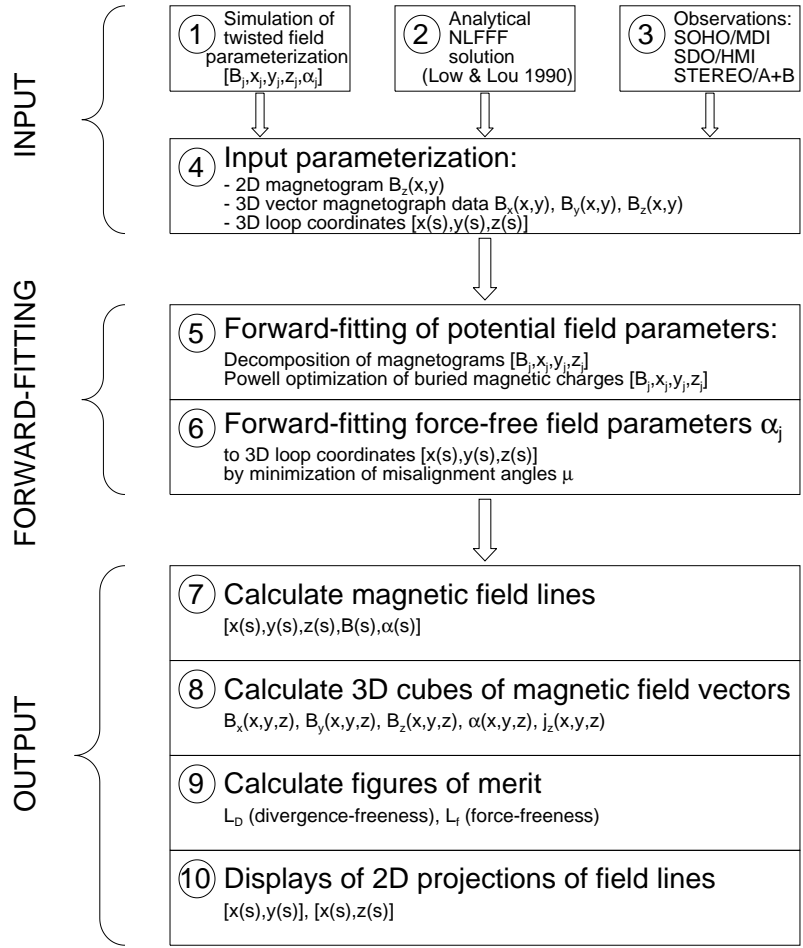}}
\caption{A flow chart of 10 modules of the forward-fitting code that
calculates nonlinear force-free field solutions from various forms
of inputs (simulations, analytical solutions, observational data).
The 10 modules are described in Section 2.}
\end{figure}

\section{	Numeric Code  			          }

A scheme of the numeric code that performs forward-fitting of nonlinear
force-free magnetic fields (NLFFF) is shown in Figure 1. The ten different
modules of the algorithm can be organized into three groups: Input
modules (1-4), forward-fitting modules (5-6), and output 
modules (7-10), which we will describe in some more detail in the
following.

\medskip\noindent
\underbar{\bf (1) Simulated Input:} This module serves to create test cases 
and defines a 3D magnetic field model directly by $n=5 N_{\rm m}$ free parameters, 
which includes the surface magnetic field strength $B_j$ and subphotospheric 
position $(x_j, y_j, z_j)$ of the buried magnetic charges, as well as the 
force-free
parameters $\alpha_j$ of the twisted magnetic field for every magnetic
charge $j=1,...,N_{\rm m}$ (see definitions in Paper I). We will use models with 
$N_{\rm m}=1-10$ magnetic charges, so we deal with $n=5-50$ input 
parameters per test case. Our models will use unipolar ($N_{\rm m}=1$), dipolar 
($N_{\rm m}=2$), quadrupolar ($N_{\rm m}=4$), and random distributions of $N_{\rm m}=10$ 
magnetic charges, where the models with multiple charges are grouped into 
pairs of opposite magnetic polarity with identical force-free parameters 
$\alpha_j=\alpha_{j+1}$ for pairs with conjugate magnetic polarization 
(to mimic a nearly force-free field). The purpose of this simulation
module is mostly to test the convergence of the code (with a large number
of free parameters), so that the output can be compared with a known input, 
regardless of other problems, such as the suitability of our parameterization 
(which is unknown for external analytical or observational data) or the 
fulfillment of the divergence-free and force-free conditions (that define 
a NLFFF solution).

\medskip\noindent
\underbar{\bf (2) Analytical NLFFF Solutions:} This module accesses 
external magnetic field data (in form of 3D cubes of magnetic field vectors)
and extrapolated field lines (which serve as proxy for coronal loops) from
a known analytical NLFFF solution. In our tests described here we will use 
solutions of a particular NLFFF model described in Low and Lou (1990),
which is also summarized and used in Malanushenko \etal (2009; Appendix A).
The Low and Lou field depends on two free parameters in the Grad-Shafranov
equation, which contains a constant $a$ and the harmonic number $n$ of the
Legendre polynomial. We will use a model with $[a=0.6, n=2.0]$,
which are also rendered in Malanushenko \etal (2009). 
Since the Low and Lou model
represents an exact analytical solution, we can test whether our code is
capable to retrieve the correct force-free parameters $\alpha({\bf x})$
in the 3D cube, as well as along individual loops, $\alpha(s)$. 
Furthermore, it will reveal whether our choice of magnetic field
parameterization $(B_j, x_j, y_j, z_j, \alpha_j)$ is suitable to represent
this particular NLFFF magnetic field, whether the forward-fitting code 
converges to the correct solution, and how divergence-free and force-free 
our analytical approximation of second order is compared with an exact
NLFFF solution. 

\medskip\noindent
\underbar{\bf (3) Observational Data Input:} This module inputs
external data directly, such as line-of-sight magnetograms $B_z(x,y)$
from SOHO/MDI or SDO/HMI, or alternatively vector fields
$[B_x(x,y), B_y(x,y), B_z(x,y)]$ if available. In addition, 
constraints on coronal field lines can be obtained from stereoscopic
triangulation from STEREO/A and B (\eg Feng \etal, 2007a; Aschwanden 
\etal, 2008a), in form of 3D field line coordinates $[x(s), y(s), z(s)]$,
where $s$ is a field line coordinate that extends from one loop footpoint
$s=0$ to the other loop footpoint at $s=L$, or to an open-field boundary
of the 3D computation box. For future applications we envision also
modeling with (automated) 2D loop tracings alone (\eg from SOHO/EIT, 
TRACE, {\it Hinode}/EIS, or SDO/AIA), without the necessity of STEREO
observations. However, 2D loop tracings represent weaker constraints
than 3D loop triangulations, and thus may imply larger ambiguities
in the NLFFF forward-fitting solution.

\medskip\noindent
\underbar{\bf (4) Input Coordinate System:} After we get input from
one of the three options (Figure 1 top), we need to bring the input data
into the same self-consistent coordinate system. Since magnetograms
are measured in the photosphere, the curvature of the solar surface has
to be taken into account. If a longitudinal magnetic field strength
$B_z(x,y)$ is measured at image position $(x,y)$, the corresponding
line-of-sight cordinate $z$ is defined by $x^2+y^2+z^2=R_{\odot}^2$, 
which defines
the 3D position of the magnetic field, $B_z(x,y,z)$. No correction of
the coordinates of the magnetogram is needed for simulated and observed 
input data. However, the analytical NLFFF solution of Low and Lou (1990)
neglects the curvature of the solar surface and yields the 3D magnetic
field vectors ${\bf B}({\bf x})$ in a cartesian grid. Hence we place the
cartesian Low and Lou solution tangentially to the solar surface and
extrapolate the magnetic field vectors to the exact position of the 
curved (photospheric) solar surface (assuming an $r^{-2}$-dependence).
After we transformed all input into the same coordinate system, normalized to
length units of solar radii ($R_{\odot}=1$) from Sun center $[0,0]$, 
we have magnetograms in 
form of $B_z(x,y,z_{\rm ph})$, or vector magnetograph data in form of 
$[B_x(x,y,z_{\rm ph}), B_y(x,y,z_{\rm ph}), B_z(x,y,z_{\rm ph})]$, with the
photospheric level at $z_{\rm ph}=\sqrt{1-x^2-y^2}$, and coronal loops
in 3D coordinates of $[x(s), y(s), z(s)]$, with $0<s<L$, and $L$ being
the length of a loop, or a segment of it. 

\medskip\noindent
\underbar{\bf (5) Forward-Fitting of Potential-Field Parameters:} 
We decompose first the line-of-sight magnetogram $B_z(x,y,z_{\rm ph})$
into a number of $N_{\rm m}$ buried magnetic charges, which produce 2D gaussian-like
local distributions $B_z(x,y)$ in the magnetogram, which are iteratively
subtracted, while the maximum field strength $B_j$ and 3D position 
$(x_j, y_j, z_j)$ is measured for each component. An early approxmiate
algorithm is shown with tests in Aschwanden and Sandman (2010;
Equation (13) and Figure 3 therein). A more accurate inversion for the
deconvolution of magnetic charges from a line-of-sight magnetogram is
derived in Aschwanden \etal (2012a; Appendix A and Figure 4 therein).
In order to obtain the maximum accuracy of this inversion, our code
used the parameters $(B_j, x_j, y_j, z_j$) of the direct inversion
as an initial guess and executes an additional forward-fitting
optimization with the Powell method (Press \etal, 1986), where
each of the $N_{\rm m}$ components is optimized by fitting the local magnetogram,
repeated with four iterations for all magnetic sources. We found that
the parameters converge already at the second iteration, given the
relatively high accuracy of the initial guess. With this step we have 
already determined 80\% of the $n=5 N_{\rm m}$ free parameters $(B_j, x_j, y_j, z_j),
\alpha_j$, $j=1,...,N_{\rm m}$, leaving only the force-free parameters 
$\alpha_j$ to be determined. If we set $\alpha_j=0$, we have already
an exact parameterization of the 3D potential field ${\bf B}_{\rm pot}({\bf x})$, 
which also predicts the transverse field components $B_x(x,y,z_{\rm ph})$ and
$B_y(x,y,z_{\rm ph})$ from the line-of-sight magnetogram $B_z(x,y,z_{\rm ph})$.

\begin{figure}
\centerline{\includegraphics[width=1.0\textwidth]{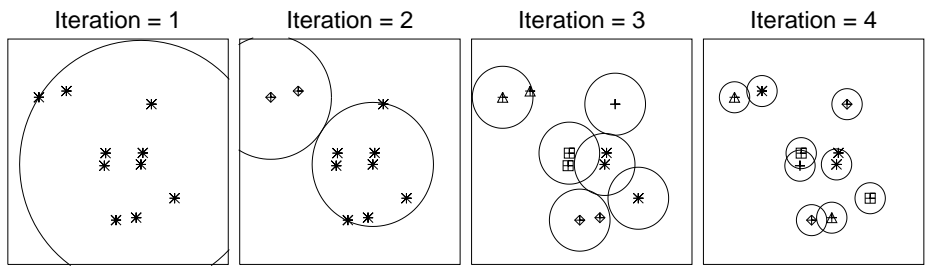}}
\caption{The scheme of hierarchical subdivision of $\alpha$-zones
(with a common force-free parameter $\alpha$) is illustrated for four
iteration cycles and $N_{\rm m}=10$ magnetic charges. 
The number of $\alpha$-zones increases with $2^{(i-1)}$ and
the radius of an $\alpha$-zone decreases with a factor $2^{(1-i)}$ in
subsequent iterations $i=1,...,4$. The number of $\alpha$-zones becomes
identical with the number of magnetic charges $j=1,...,N_{\rm m}$ after
four iteration cycles. This number of free parameters $\alpha_i$ to be
optimized is this way is reduced to 1, 2, 6, and 9 in subsequent iteration 
cycles for this example.}
\end{figure}

\medskip\noindent
\underbar{\bf (6) Forward-Fitting of Non-Potential-Field Parameters:} 
For the forward-fitting of the force-free parameters $\alpha_j$ for each
magnetic charge $j=1,...,N_{\rm m}$ we can use either the constraints of the
coronal loops ($q_{\rm v}=0$), or the transverse components of the vector-magnetograph
data ($q_{\rm v}=1$), or a combination of both ($0<q_{\rm v}<1$), which we select with 
a weighting factor $q_{\rm v}$ in the optimization of the overall misalignment 
angle $\mu$, \ie 
\begin{equation}
	\mu = q_{\rm v} \mu_{\rm loop} + (1-q_{\rm v}) \mu_{\rm vect} \ .
\end{equation}
The forward-fitting of the best-fit force-free parameters $\alpha_j$ 
is performed by iterating the calculation of the 3D misalignment angle,
which is defined for loops (or equivalently for a vector-magnetograph
3D field vector) by,
\begin{equation}
	\mu_{\rm loop} = \cos^{-1}\left(
		{ {\bf B}^{\rm theo} \cdot {\bf B}^{\rm obs}) \over
		 |{\bf B}^{\rm theo}| \cdot |{\bf B}^{\rm obs}|} \right) \ ,
\end{equation}
between the theoretically 
calculated loop field lines ${\bf B}^{\rm theo}$ based on a trial set of
parameters $(B_j, x_j, y_j, z_j), \alpha_j$, $j=1,...,N_{\rm m}$, and the
observed field direction ${\bf B}^{\rm obs}$ of the observed loops. The overall
misalignment angle is averaged (quadratically) from $N_{\rm seg}=10$ loop positions
in all $N_{\rm loop}$ loops. The variation of the trial sets of $\alpha_j$ is
accomplished by a progressive subdivision of magnetic zones in subsequent
iterations, starting from a single value for the entire active region
(which corresponds to a linear force-free field model), and
progressing with zones that become successively smaller by a factor of
$2^{i-1}$, with $i=1,...,N_{\rm iter}$ the number of iterations. The hierarchical
subdivision of $\alpha$-zones procedes in order of decreasing magnetic
field strength $B_j$. In each iteration all magnetic zones 
are successively varied, and for each zone the force-free parameter
$\alpha_j$ is varied within a range of $|\alpha_j| < \alpha_{\rm max}$, until
a minimum of the overall misalignment angle $\mu$ is found. An example
of a hierarchical subdivision of $\alpha$-zones in subsequent iterations
is shown in Figure 2. For the test images we have chosen a dimension of 
$N_x=N_y=60$, for which 
the subdivision of zone radii reaches a lower limit of one pixel after about 
five iterations (since $2^5=32 \approx N_x/2$). Thus, after five iterations,
all magnetic sources are fitted individually in each iteration step. 
Convergence is generally reached for $N^{\rm iter} \lapprox 10-20$ iteration 
cycles. The computation scales linearly with the number $N_{\rm m}$ of magnetic
sources and the number $N_{\rm loop}$ of fitted loops.

\medskip\noindent
\underbar{\bf (7) Calculating Magnetic Field Lines:} 
Once our forward-fitting algo\-rithm converged and determined a full set
of $n=5 N_{\rm m}$ free parameters, $(B_j, x_j, y_j, z_j,$ $\alpha_j)$, 
$j=1,...,N_{\rm m}$, we can calculate the magnetic field vector ${\bf B}({\bf x})$
of the quasi-forcefree field at any arbitrary location ${\bf x}=(x,y,z)$
in space (see Equations (34)--(42) in Paper I). To calculate the magnetic field
along a particular field line $[x(s), y(s), z(s)]$, we just step
iteratively by increments $\Delta s$,
\begin{equation}
        \begin{array}{ll}
                x(s+\Delta s) &= x(s) + \Delta s [B_x(s)/B(s)] p \\
                y(s+\Delta s) &= y(s) + \Delta s [B_y(s)/B(s)] p \\
                z(s+\Delta s) &= z(s) + \Delta s [B_z(s)/B(s)] p \\
        \end{array} \ .
\end{equation}
where $p=\pm1$ represents the sign or polarzation of the magnetic charge,
and thus can be flipped to calculate a field line into opposite direction.

\medskip\noindent
\underbar{\bf (8) Calculation of 3D Data Cubes:} 
By the same token we calculate 3D cu\-bes of magnetic field vectors
${\bf B}({\bf x})=B_x(x_i,y_j,z_k), B_y(x_i,y_j,z_k), 
B_z(x_i,y_j,z_k)$, in a cartesian grid $(i,j,k)$ with $i=1,...,N_x$,
$j=1,...,N_y$, $k=1,...,N_z$. The 3D cubes of force-free parameters
$\alpha(x_i, y_j, z_k)$ can be calculated from the $B(x_i, y_j, z_k)$
cubes, for each of the three vector components,
\begin{equation}
        \alpha_x({\bf x}) = {1 \over 4 \pi}
        {(\nabla \times {\bf B})_x \over {\bf B}_x} 
	= {1 \over 4\pi B_x} \left({\partial B_z \over \partial y}
             - {\partial B_y \over \partial z}\right) \ ,
\end{equation}
\begin{equation}
        \alpha_y({\bf x}) = {1 \over 4 \pi}
        {(\nabla \times {\bf B})_y \over {\bf B}_y} 
	= {1 \over 4\pi B_y} \left({\partial B_x \over \partial z}
             - {\partial B_z \over \partial x}\right) \ ,
\end{equation}
\begin{equation}
        \alpha_z({\bf x}) = {1 \over 4 \pi}
        {(\nabla \times {\bf B})_z \over {\bf B}_z} 
	= {1 \over 4\pi B_z} \left({\partial B_y \over \partial x}
             - {\partial B_x \over \partial y}\right) \ .
\end{equation}
using a second-order scheme to compute the spatial derivatives,
\ie $\partial B_x / \partial y = (B_{i+1j,k}-B_{i-1,j,k})
/2 (y_{i+1}-y_{i-1})$. In principle, the three values $\alpha_x$,
$\alpha_y$, $\alpha_z$ should be identical, but the numerical 
accuracy using a second-order differentiation scheme is most
handicapped for those loop segments with the smallest values of
the $B$-component (appearing in the denominator), for instance
in the $\alpha_z$ component $\propto (1/B_z)$ near the loop tops
(where $B_z \approx 0$). It is therefore most advantageous to use
all three parameters $\alpha_x$, $\alpha_y$, and $\alpha_z$ in 
a weighted mean, 
\begin{equation}
	\alpha = {\alpha_x w_x + \alpha_y w_y + \alpha_z w_z 
	\over {w_x + w_y + w_z}} \ ,
\end{equation}
but weight them by the magnitude of the (squared) magnetic field 
strength in each component, 
\begin{equation}
	w_x = {B_x^2} \ , \quad 
	w_y = {B_y^2} \ , \quad 
	w_z = {B_z^2} \ ,  
\end{equation}
so that those segments have no weight where the $B$-component
approaches zero. With this method, we can determine the force-free parameter
$\alpha(x_i, y_j, z_k)$ at any given 3D grid point $[x_i, y_j, z_k]$,
as well as along a loop coordinate, $\alpha(s)$. 

The 3D cubes of current densities
${\bf j}=(j_x, j_y, j_z)$ follow from the
definition ${\bf j}/c = (\nabla \times {\bf B})/(4\pi) =
\alpha({\bf x}) {\bf B}$,
\begin{equation}
	{\bf j}(x_i, y_j, z_k) = c \ \alpha(x_i, y_j, z_k) 
	{\bf B}(x_i, y_j, z_k) \ .
\end{equation}

\medskip\noindent
\underbar{\bf (9) Calculation of Figures of Merit:} 
Figures of merit (how physical a converged NLFFF solution is) can be computed
for the divergence-freeness $\nabla \cdot {\bf B} = 0$ compared to the
field gradient $B/\Delta x$ over a pixel length $\Delta x$,
\begin{equation}
        L_{\rm d} = {1 \over V} \int_V
        {|(\nabla \cdot {\bf B}) |^2
        \over |B / \Delta x|^2} dV \ .
\end{equation}
Similarly, the force-freeness can be quantified by the ratio of the
Lorentz force, $({\bf j} \times {\bf B}) \propto (\nabla \times {\bf B}) \times
{\bf B}$ to the normalization constant $B^2 / \Delta x$,
\begin{equation}
        L_{\rm f} = {1 \over V} \int_V
        {|(\nabla \times {\bf B}) \times {\bf B}|^2
        \over |B^2 / \Delta x|^2}  dV \ ,
\end{equation}
where $B = |{\bf B}|$. We calcuate these quantities in agreement with
the definitions given in Paper I. 

\medskip\noindent
\underbar{\bf (10) Display of 2D Projections:} 
For visualization purposes of the 3D field, of both the numerically
calculated solution (of our quasi-NLFFF model) as well as for the
observed loops, it is most practical to display the field lines
in the three orthogonal projections, \ie
$[x(s), y(s)]$ for a top-down view, or 
$[x(s), z(s)]$ and $[y(s), z(s)]$ for side views.  

\begin{table}
\caption{Standard control parameter settings of the forward-fitting code
used in the tests of this study.}
\begin{tabular}{ll}
\hline
Parameter & Description \\
\hline
$N_{\rm grid}=8$    &Grid size in pixels for loop footpoint selection \\
$\Delta x=0.0034$ &Pixel size of computation grid (in solar radii) \\
${\rm Thresh}=0$     &Threshold of magnetic field [gauss] for loop footpoint selection \\
$N_{\rm mag}=10$	&Maximum number of magnetic charges   \\
$q_{\rm mag}=0.001$ &Residual limit $B/B_{\rm max}$ of magnetogram decomposition \\
nsm=0		&Smoothing of magnetogram (in number of boxcar pixels) \\
$i_{\rm opt}=4$     &Number of cycles for optimization of potential field parameters\\
Meth=A 	&Method of subdividing magnetic zones \\
$N_{\rm iter}=20$   &Maximum number of iteration cycles   \\
$\Delta s=\Delta x$ &Spatial resolution along field line (in solar radii) \\
$N_{\rm seg}=10$   &Number of loop segments for misalignment angle calculation\\
$h_{\rm max}=3.5 \Delta x$ &Maximum altitude range for magnetogram calculation
		(solar radii) \\
$h_{\rm alt}=0.15$  &Maximum altitude range for field line extrapolation \\
$\alpha_{\rm max}=100.$ &Maximum range for force-free $\alpha$ per iteration
		(solar radius$^{-1}$) \\
$acc=0.001$     &Relative accuracy in $\alpha$ optimization step \\
$q_{\rm loop}=0.5$  &Relative loop position for starting of field line computation\\
$q_{\rm zone}=0.5$  &Magnetic zone diminuishing factor in subsequent iterations\\
$q_{\rm v}=0.0$       &Weighting factor of loop data vs. vector magnetograph data \\
eps=0.1       &Convergence criterion for change in misalignent angle (deg) \\
\hline
\end{tabular}
\end{table}

\medskip\noindent
\underbar{\bf Control Parameter Settings:}
The numeric forward-fitting code has a number of control parameter
settings, which can be changed individually to optimize the performance 
or the computation speed of the code. 
We list the set of standard control parameter settings in Table 1,
which are generally used in this Paper if not mentioned otherwise.
These parameters control: the selection of loop field lines 
(module 1-3: $N_{\rm grid}$, $\Delta x$, Thresh), the decompostion of the 
magnetogram (module 4: $N_{\rm mag}$, $q_{\rm mag}$, nsm, $i_{\rm opt}$), and
the forward-fitting of the force-free $\alpha$ parameter (module 5: 
Meth, $N_{\rm iter}$, $\Delta s$, $N_{\rm seg}$, $h_{\rm max}$, $h_{\rm alt}$, 
$\alpha_{\rm max}$, acc, $q_{\rm loop}$, $q_{\rm zone}$, $q_{\rm v}$, eps). 

\begin{figure}
\centerline{\includegraphics[width=1.0\textwidth]{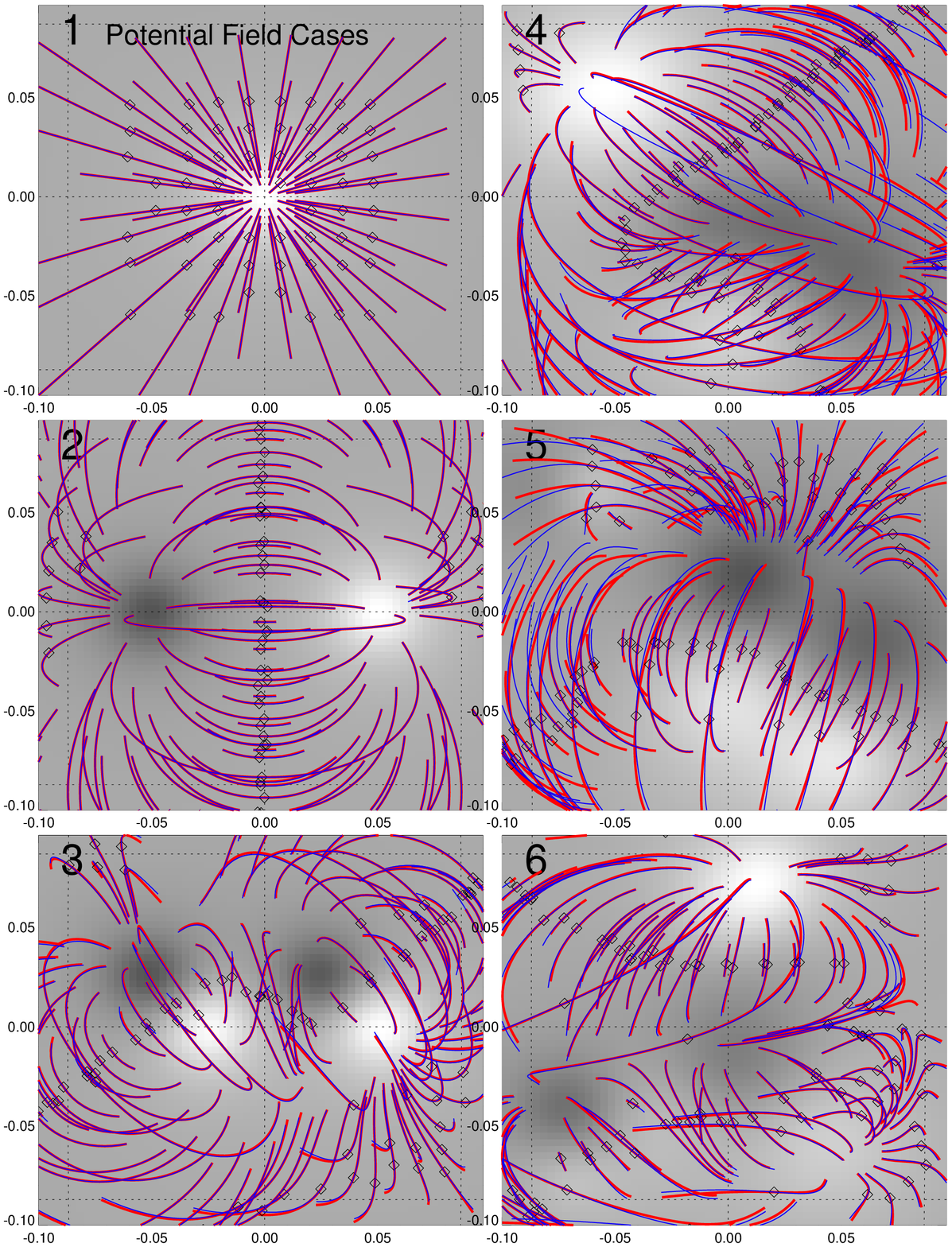}}
\caption{Test cases \# 1-6 are shown, consisting of a unipolar 
(\#1: top left), a dipolar (\#2: middle left), a quadrupolar
(\#3: bottom left), and three decapolar cases (\#4-6: panels on right side).
The displays contain the line-of-sight magnetograms (greyscale),
the theoretically simulated loop field lines (red curves), and the
overlaid best-fit NLFFF field lines (blue curves). The starting point
of the calculated field lines are indicated with diamonds (at midpoint
of loops, $q_{\rm loop}=0.5$). Note the small amount of misalignment,
ranging from $\mu=0.0^\circ$ (\# 1) to $\mu=6.6^\circ$ (\#5) (the 
values are given in Table 2).}
\end{figure}

\begin{table}
\caption{Best-fit parameters of forward-fitting of the NLFFF model to potential 
field cases (with $\alpha_j=0$), using standard settings of the forward-fitting 
code (Table 1). The columns contain the case \#=1-6, the number of magnetic
charges $N_{\rm mag}$, the number of loop field lines $N_{\rm loop}$, the mean
misalignment angle $\mu$, the mean best-fit force-free parameter $\alpha$ 
per loop, the divergence-freeness figure of merit $L_{\rm d}$, the force-freeness
figure of merit $L_{\rm f}$, and the computation time $t_{\rm CPU}$ of the forward-fitting
module 6. The last lines of the Table contain the means and
standard deviations $\sigma$ of the six cases.}
\begin{tabular}{rrrrrrrr}
\hline
\# &$N_{\rm mag}$ & $N_{\rm loop}$ & $\mu$ & $\alpha$ & $L_{\rm d}$ & $L_{\rm f}$ & $t_{\rm CPU}$ \\
\hline
1  &  1 &  61 & $0.0^\circ$ &$ 0.00$ & 0.000001 & 0.000001 &   2 s \\
2  &  2 &  91 & $2.7^\circ$ &$ 0.00$ & 0.000002 & 0.000002 &   9 s \\
3  &  4 &  91 & $3.8^\circ$ &$-0.03$ & 0.000007 & 0.000008 &  23 s \\ 
4  & 10 & 107 & $3.0^\circ$ &$-0.06$ & 0.000001 & 0.000008 & 111 s \\
5  & 10 &  95 & $6.6^\circ$ &$-0.08$ & 0.000017 & 0.000480 & 117 s \\
6  & 10 &  98 & $4.2^\circ$ &$-0.39$ & 0.000003 & 0.000003 & 105 s \\
\hline
Mean &   &    & $3.4^\circ$    &$-0.09$   &     0.000005  &      0.000084 &      61  s \\
$\pm\sigma$&& & $\pm2.1^\circ$ &$\pm0.15$ & $\pm0.000006$ & $\pm$0.000194 & $\pm 55$ s \\
\hline
\end{tabular}
\end{table}

\section{	Potential Field Tests  		          }

A first set of six test cases consists of potential field models
(with $\alpha_j=0$), including a unipolar charge, a dipole,
a quadrupole, and three cases with 10 randomly distributed magnetic
sources, identical to Cases \#1-3 in Paper I, and to Cases \#7-9 (but
with $\alpha_j$ set to zero). For each of these six cases we show in Figure 3 
a set of field lines calculated from the model (Figure 3, red curves),
and a set of field lines obtained from forward-fitting with our
NLFFF code. The agreement between the two sets of field lines can
be expressed by the mean 3D misalignment angle $\mu$ (Equation (2)),
which is found to be very small, within a range of 
$\mu=0.0^\circ-6.6^\circ$, or $\mu=3.4^\circ\pm2.1^\circ$. 
The individual values are listed in Table 2 (forth column). 
While this mostly represents a test of the accuracy of
module 5 (forward-fitting of potential-field parameters), the algorithm 
treats the force-free parameter $\alpha_j$ as a variable too, and thus
it represents also a test of the accuracy in determining this parameter
in general. Compared with the theoretical value as it was set in the
simulation of the input magnetogram ($\alpha_j=0$),
the best-fit values are found to be $\alpha=-0.09\pm0.15$ (Table 2,
fifth column), which corresponds to 
$\Delta N_{\rm twist}=b l / 2 \pi=\alpha l / 4 \pi = \pm 0.0018$ 
twist turns over the length $l=0.05 \pi =0.157$ solar radii
of a typical field line (see definitions in Equations (16)--(17) in Paper I). 
Thus the uncertainty of our forward-fitting corresponds to less than 
$\pm0.2$\% of a full twist turn over a loop length. Another measure
of the quality of the NLFFF forward-fit is the divergence-freeness,
which is found to be $L_{\rm d} = (5 \pm 6) \times 10^{-6}$ (Table 2, sixth column), 
and the force-freeness, which is found to be $L_{\rm f}=(84\pm 194) \times 10^{-6}$ 
(Table 2, seventh column), both being extremely accurate.
The average computation time for the NLFFF forward-fitting runs of potential
field cases was found to be $t_{\rm CPU} \approx 61$ s (on a Mac OS X with
$2 \times 3.2$ GHz Quad-Core Intel Xeon processor and 32 GB 800 MHz
DDR2 FB-DIMM Memory).  

\begin{table}
\caption{Best-fit parameters of forward-fitting of the NLFFF model to potential 
field cases (with $\alpha_j=0$), using some non-standard settings in
the spatial resolution $\Delta s/\Delta x$ of calculated field lines,
the number of magnetic source compoonents $N_{\rm mag}$, 
the starting point of field line extrapolation $q_{\rm loop}$,
the relative weighting of loop and vector magnetograph data $q_{\rm v}$,
but otherwise standard settings as listed in Table 1.}
\footnotesize
\begin{tabular}{llllrrrr}
\hline
$\Delta s$ & $N_{\rm mag}$ & $q_{\rm loop}$ & $q_{\rm v}$ & $\mu$ & $\alpha$ & $L_{\rm d} [10^{-6}]$ & $t_{\rm CPU}$ \\
\hline
$\times 1.0$ & $\times 1$ & 1.0 & 0.0 & $3.4^\circ\pm2.1^\circ$ & $-0.09\pm0.15$ & $5\pm6$ &  $61\pm55$ s \\
$\times 0.5$ & $\times 1$ & 1.0 & 0.0 & $3.4^\circ\pm2.1^\circ$ & $-0.08\pm0.14$ & $5\pm6$ &  $70\pm66$ s \\
$\times 1.0$ & $\times 2$ & 1.0 & 0.0 & $3.2^\circ\pm1.7^\circ$ & $ 0.03\pm0.13$ & $10\pm19$& $228\pm233$ s \\
$\times 1.0$ & $\times 1$ & 0.0 & 0.0 & $3.4^\circ\pm2.1^\circ$ & $-0.07\pm0.16$ & $5\pm6$ &  $71\pm66$ s \\
$\times 1.0$ & $\times 1$ & 1.0 & 1.0 & $1.8^\circ\pm2.3^\circ$ & $-0.07\pm0.10$ & $3\pm3$ &  $314\pm290$ s \\
\hline
\end{tabular}
\end{table}

We performed also some parametric studies to explore the accuracy of the
forward-fitting code as a function of some control parameters that
are different from the standard settings given in Table 1. We list
the results in Table 3. If we increase the spatial resolution of the
field line extrapolation to $\Delta s/\Delta x=0.5$, the
accuracy of the field lines does not change, neither in terms of the
the mean misalignment angle nor in the divergence-freeness figure
of merit (Table 3, second line). Increasing the number of magnetic
components in the decomposition of magnetograms does not improve the
accuracy for the potential-field cases (\eg by a factor of two compared
with the simulated numbers of $N_{\rm mag}=1,2,4,10$), but degrades the
divergence-freeness and force-freeness and increases the computation
time by a factor of $\approx 4$ (Table 3, third line). 
Starting the field line extrapolation
at the footpoints ($q_{\rm loop}=0.0$), rather than from the loop midpoints
($q_{\rm loop}=0.5$), leads to no significant improvement (Table 3, fourth line).
Changing the weighting of coronal loop constraints ($q_{\rm v}=0$) to
using only photospheric vector magnetograph data ($q_{\rm v}=1$)
improves the misalignment to $\mu=1.8^\circ \pm 2.3^\circ$, which
represents an improvement in the accuracy by about a factor of two,
but requires about five times more computation time. 
Thus, the accuracy in fitting potential field cases is fairly robust
and does not depend the detailed setting of control parameters,
except for the weigthing of photospheric versus coronal constraints.

\begin{figure}
\centerline{\includegraphics[width=1.0\textwidth]{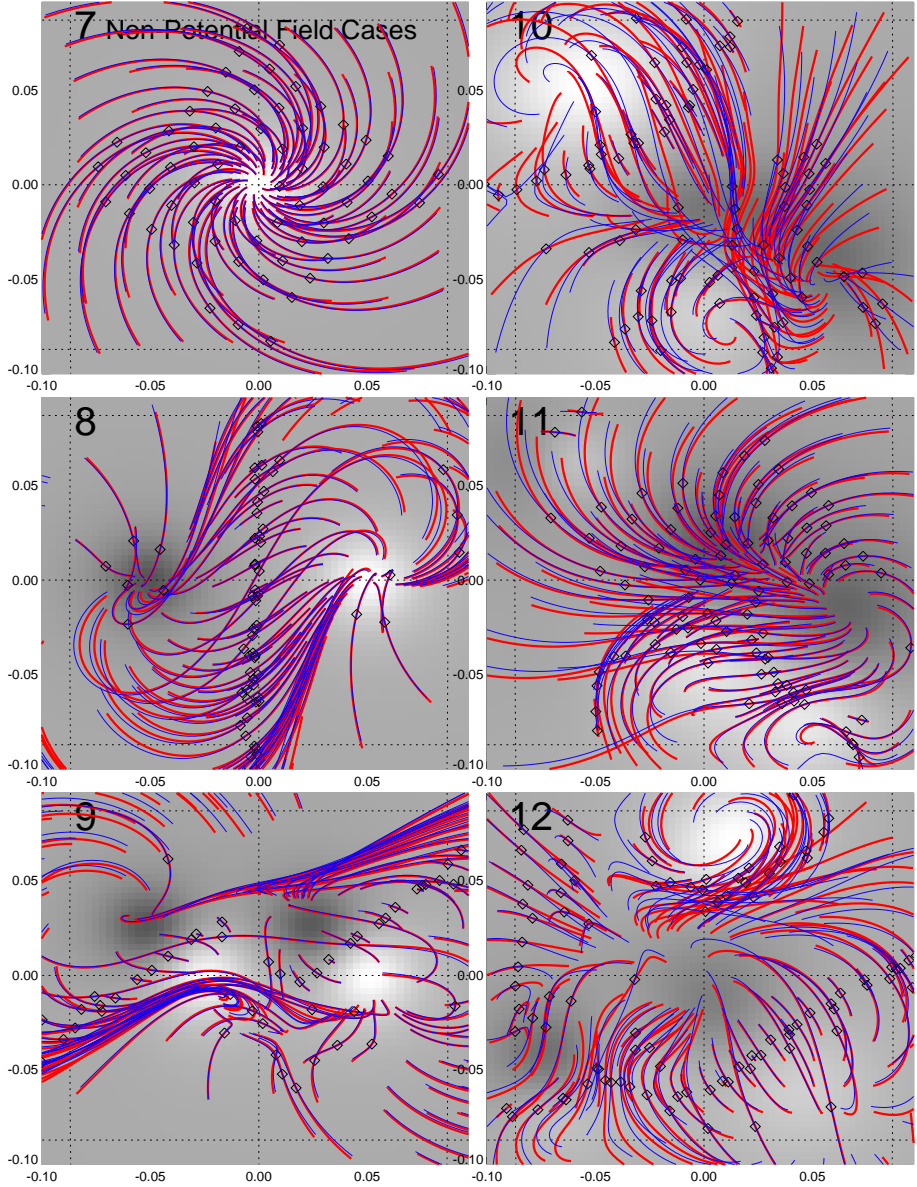}}
\caption{Test cases \# 7-12 are shown, consisting of a unipolar 
(\#7: top left), a dipolar (\#8: middle left), a quadrupolar
(\#9: bottom left), and three decapolar cases (\#10-12: panels on right side).
The displays contain the line-of-sight magnetograms (greyscale),
the theoretically simulated loop field lines (red curves), and the
overlaid best-fit NLFFF field lines (blue curves). The starting points
of the calculated field lines are indicated with diamonds (at midpoint
of loops, $q_{\rm loop}=0.5$). The misalignment angles between the theoretical
models and the best fits are listed in Table 4. Note the huge difference of
field line topologies compared with the potential-field cases (shown
in Figure 3), although the line-of-sight magnetograms are identical.}
\end{figure}

\section{	Forward-Fitting to Quasi-NLFFF Models      }

Now we present the first tests of forward-fitting to non-potential fields
(with $\alpha_j \ne 0$), numbered as test cases \# 7-12. These six cases
have the same line-of-sight magnetograms $B_z(x,y)$ or magnetic charges
$(B_j, x_j, y_j, z_j)$ as the potential-field cases \# 1-6, but have
a different twist or force-free parameter $\alpha_j$. We show the
magnetograms and the theoretical field lines of the models in Figure 4 
(red curves), and the best-fit field lines of our NLFFF forward-fitting code 
in Figure 4 (blue curves), using standard control parameter settings (Table 1).
The misalignment between these two sets of simulated and
forward-fitted field lines amounts to $\mu=0.7^\circ-12.8^\circ$, or
$\mu=5.1^\circ\pm4.3^\circ$ (Table 4, fourth column). 
These test results are quite satisfactory, first of all since the difference
between the theoretical and best-fit field lines in Figure 4 are hardly
recognizable by eye, and thus will suffice for all practical purposes,
and secondly, the misalignment is about an order of magnitude smaller
than found between traditional NLFFF codes and stereoscopically triangulated
coronal loops ($\mu \approx 24^\circ-44^\circ$; DeRosa \etal, 2009).
We see that the force-free parameters vary substantially, in a range of 
$\alpha = 6 \pm 40$ (solar radius$^{-1}$) (Table 4, 5th column), 
which translates into a number  
$N_{\rm twist} = \alpha l / 4 \pi \approx 0.5$ of (full) twist turns 
over a typical loop length. The merit of figure for the divergence-freeness 
is $L_{\rm d}=(0.8 \pm 0.5) \times 10^{-3}$ (Table 4, sixth column), and the merit 
of figure for the force-freeness is $L_{\rm f}=(2.3 \pm 2.3) \times 10^{-3}$ 
(Table 4, seventh column).
The computation time is ($t_{\rm CPU} \approx 100$ s) less than a factor of two 
longer than for the potential-field cases (Table 2).

\begin{table}
\caption{Best-fit parameters of forward-fitting of the NLFFF model to 
non-potential field cases (with $\alpha_j \ne 0$), using standard 
settings of the forward-fitting 
code (Table 1). The columns contain the cases \#=7-12, the number of magnetic
charges $N_{\rm mag}$, the number of loop field lines $N_{\rm loop}$, the mean
misalignment angle $\mu$, the mean input force-free $\alpha$ parameter
values, the divergence-freeness figure of merit $L_{\rm d}$, the force-freeness
figure of merit $L_{\rm f}$, and the computation time $t_{\rm CPU}$ of the forward-fitting
module 6.}
\begin{tabular}{rrrrrrrr}
\hline
\# &$N_{\rm mag}$ & $N_{\rm loop}$ & $\mu$ & $\alpha$ & $L_{\rm d}$ & $L_{\rm f}$ & $t_{\rm CPU}$ \\
\hline
7  &  1 & 66 & $0.7^\circ$ &$-20$      & 0.000453 & 0.000299 &  2 s \\
8  &  2 & 85 & $2.2^\circ$ &$-20\pm 1$ & 0.000253 & 0.000104 &  8 s \\
9  &  4 & 82 & $3.6^\circ$ &$-30\pm12$ & 0.000727 & 0.000691 & 21 s \\ 
10 & 10 & 89 & $12.8^\circ$&$ 29\pm40$ & 0.001813 & 0.004672 & 118 s \\
11 & 10 & 89 & $4.5^\circ$ &$  2\pm102$& 0.000784 & 0.003123 & 179 s \\
12 & 10 & 99 & $7.1^\circ$ &$ 74\pm62$ & 0.000976 & 0.005334 & 271 s \\
\hline
Mean &   &   & $   5.1^\circ$ & 6       &  0.000834     & 0.002370 & 99 s \\
$\pm\sigma$&&& $\pm4.3^\circ$ & $\pm40$ & $\pm0.000543$ & $\pm$0.002319 &$\pm 109$ s \\
\hline
\end{tabular}
\end{table}

In order to achieve the most accurate performance of our code we explored
also other control parameter settings than the standard parameters given
in Table 1. Instead of using the hierarchical $\alpha$-zone subdivision
as shown in Figure 2 (Meth=A), we tested also other methods, such as
subdivision by magnetically conjugate pairs of magnetic charges (Meth=B),
or subdivision by magnetically conjugate loop footpoints (Meth=C). 
In 90\% of the test cases all three methods converged to the same minimum
misalignment angle within $\pm 0.1^\circ$, but for the
10\% of discrepant cases method $A$ performed always best, so we conclude
that method $A$ is the most robust one. 

Increasing the resolution of calculating field lines to 
$\Delta s = 0.5 \Delta x$ does not improve the misalignment  
($\mu = 5.1^\circ\pm 4.3^\circ$; Table 5, second case);
Increasing the number of magnetic sources by a factor of two 
does not improve the misalignment significantly either
(Table 5; third case). Starting the field line extrapolation 
at the footpoints ($q_{\rm loop}=0.0$) rather than from the loop midpoints, 
has no effect either (Table 5; forth case). However, 
the change of replacing the coronal ($q_{\rm v}=0$) to
photospheric constraints, using 3D vector magnetograph data $q_{\rm v}=1$ 
does improve the best fits substantially, but is more costly in 
computation time (Table 5, fifth case). The reason for this improvement
is probably that photoshperic field vectors are more uniformly
distributed than coronal loops, but coronal constraints are
more important when the photospheric magnetic field is not force-free.

\begin{figure}
\centerline{\includegraphics[width=1.0\textwidth]{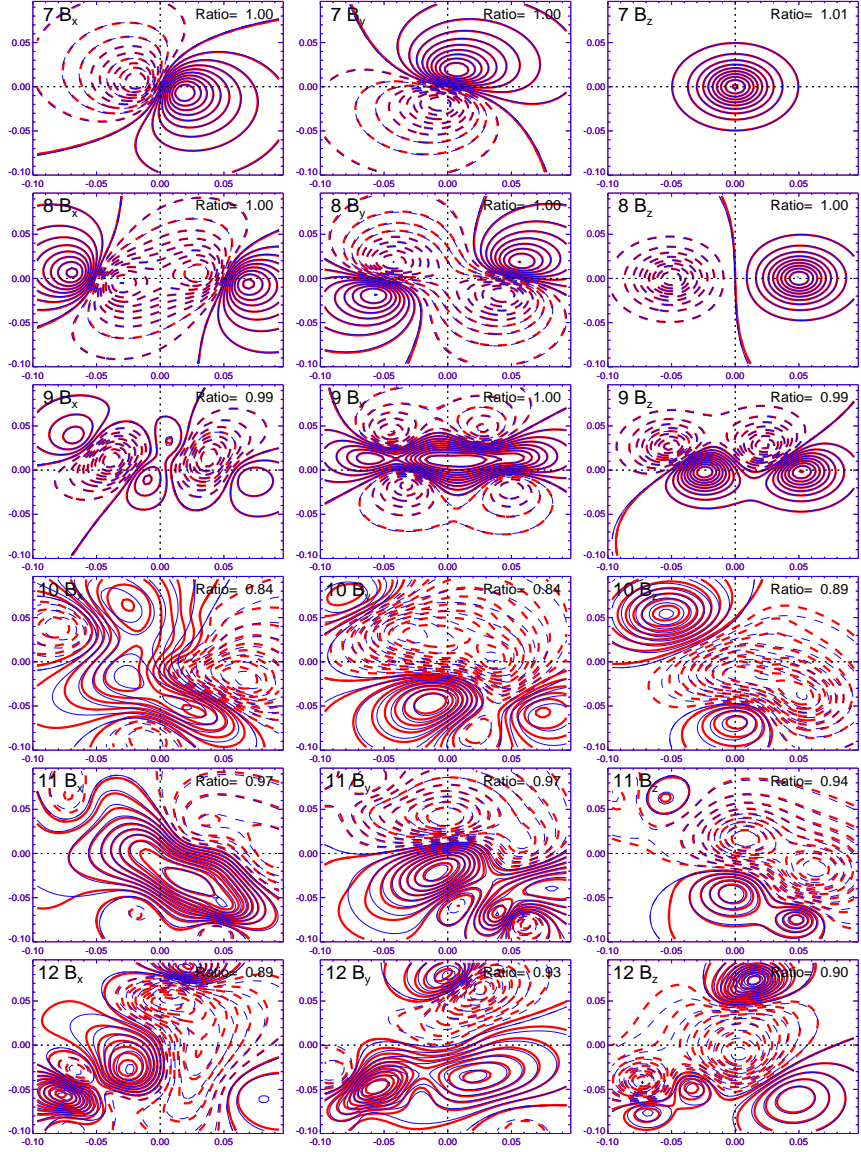}}
\caption{Contour maps of magnetic field component maps $B_x(x,y)$ (left
column), $B_y(x,y)$ (middle column), and line-of-sight component 
$B_z(x,y)$ at the photospheric level for cases \#7-12 (rows),
shown with red contours (solid for positive and dashed for negative 
magnetic polarity). The best fits that result from the decomposition
of the line-of-sight component are shown with blue curves, and the
mean ratio of the absolute magnetic field strengths between the best fit
and the model are indicated in each frame.}
\end{figure}

\begin{figure}
\centerline{\includegraphics[width=1.0\textwidth]{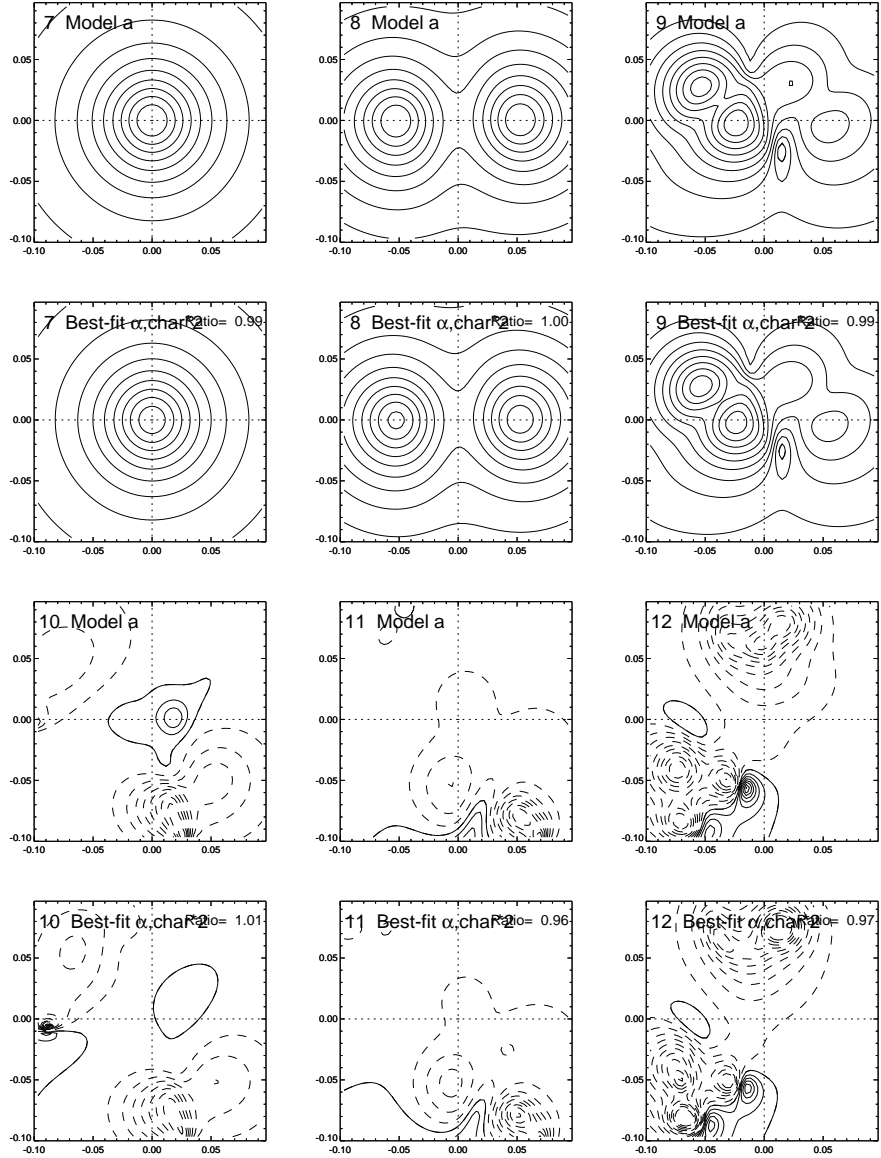}}
\caption{Contour maps of the force-free parameter $|\alpha(x,y)|$ of
the simulated models (top and third row) and best-fit solutions
(second and forth row), for the six cases \# 7-12.}
\end{figure}

\begin{table}
\caption{Best-fit parameters of forward-fitting of the NLFFF model to 
potential field cases (with $\alpha_j=0$), using some non-standard settings 
in the spatial resolution $\Delta s/\Delta x$ of calculated field lines,
the number of magnetic source compoonents $N_{\rm mag}$, 
the starting point of field line extrapolation $q_{\rm loop}$,
the relative weighting of loop and vector magnetograph data $q_{\rm v}$,
but otherwise standard settings as listed in Table 1.}
\footnotesize
\begin{tabular}{llllrrrr}
\hline
$\Delta s$ & $N_{\rm mag}$ & $q_{\rm loop}$ & $q_{\rm v}$ & $\mu$ & $L_{\rm d} [10^{-3}]$ & $L_{\rm f} [10^{-3}]$ & $t_{\rm CPU}$ \\
\hline
$\times 1.0$ & $\times 1$ & 1.0 & 0.0 & $5.1^\circ\pm4.3^\circ$ & $ 0.8\pm0.5$ & $2.3\pm2.3$ & $99\pm109$ s \\
$\times 0.5$ & $\times 1$ & 1.0 & 0.0 & $5.1^\circ\pm4.3^\circ$ & $ 0.8\pm0.6$ & $2.3\pm2.3$ & $100\pm110$ s \\
$\times 1.0$ & $\times 2$ & 1.0 & 0.0 & $4.5^\circ\pm2.9^\circ$ & $ 0.7\pm0.3$ & $2.1\pm2.1$ & $303\pm307$ s \\
$\times 1.0$ & $\times 1$ & 0.0 & 0.0 & $5.1^\circ\pm4.3^\circ$ & $ 0.8\pm0.5$ & $2.4\pm2.3$ & $100\pm110$ s \\
$\times 1.0$ & $\times 1$ & 1.0 & 1.0 & $3.4^\circ\pm3.6^\circ$ & $ 0.7\pm0.3$ & $2.2\pm2.1$ & $459\pm342$ s \\
\hline
\end{tabular}
\end{table}

The agreement between the best forward-fitting solutions of the magnetic 
field components $(B_x, B_y, B_z)$ and the model are shown in Figure 5. 
Note that only the line-of-sight magnetogram $B_z(x,y,z_{\rm ph})$ was used 
as input to the forward-fitting code, for standard control parameter settings
($q_{\rm v}=0$). For these tests, the code predicts the transverse component
maps $B_x(x,y)$ and $B_y(x,y)$, which is quite satisfactory for this
set of tests, as Figure 5 demonstrates. The mean ratios of the absolute
magnetic field strenghts are accurate within a few percents (indicated
in each panel of Figure 5). 

\begin{figure}
\centerline{\includegraphics[width=1.0\textwidth]{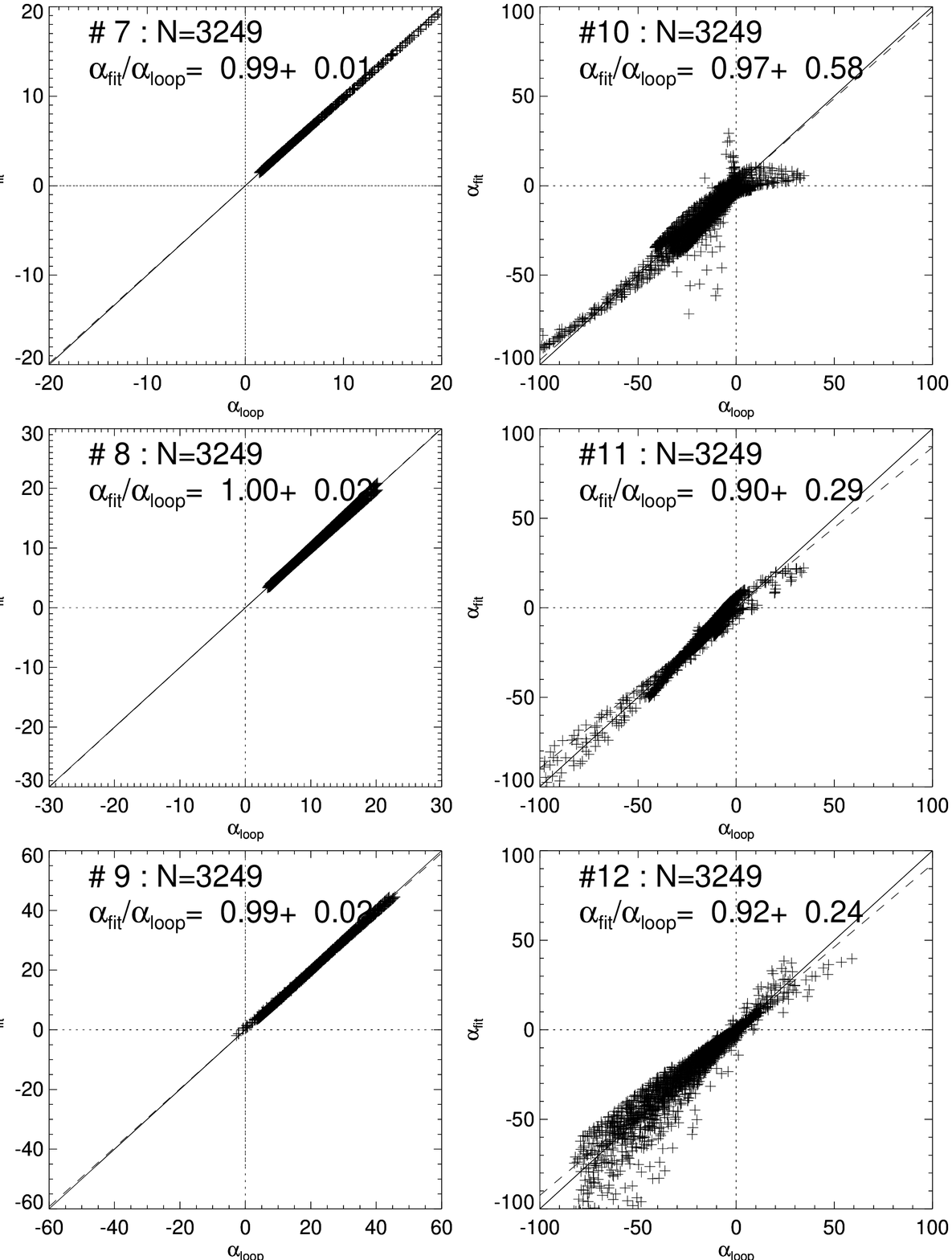}}
\caption{Scatter plot of the best-fit force-free parameters $\alpha_{\rm fit}(x,y,z_{\rm ph})$ 
of every map pixel $(x,y)$ versus the corresponding value $\alpha_{\rm sim}(x,y,z_{\rm ph})$
of the simulated models for the six cases \# 7-12. The mean and standard deviation
of the ratio $\alpha_{\rm fit}/\alpha_{\rm sim}$ is indicated in each panel.}
\end{figure}

\begin{figure}
\centerline{\includegraphics[width=1.0\textwidth]{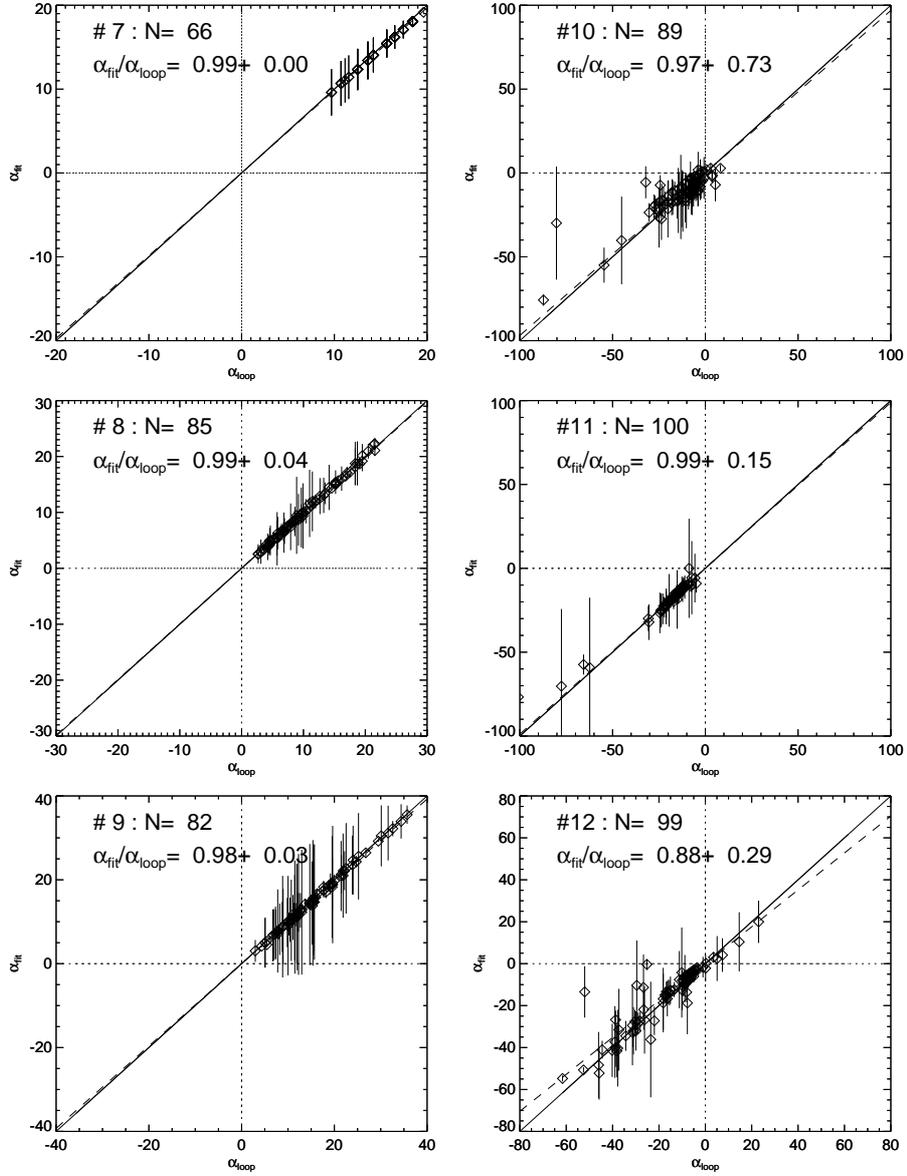}}
\caption{Scatter plot of the best-fit force-free parameters $\alpha_{\rm fit}$ 
averaged from each fitted coronal loop versus the corresponding value $\alpha_{\rm sim}$
of the simulated model loops for the six cases \# 7-12. The vertical error bars
indicate the standard deviation of the spatial variation of $\alpha_{\rm fit}(s)$ 
along each loop. The mean and standard deviation of the ratio $\alpha_{\rm fit}/\alpha_{\rm sim}$ 
is indicated in each panel.}
\end{figure}

The force-free parameter $\alpha$ is shown as a photospheric map
$|\alpha(x,y,z_{\rm ph})|$ for the model (Figure 6, top and third row) and for the
forward-fitting solution (Figure 6, second and forth row). The comparison can be
quantified by the ratio of the two values, which agrees within a
few percents. A sensible test is also to display a scatterplot of the best-fit 
$\alpha$-values versus the model $\alpha$-values for each pixel of a 
photospheric map (Figure 7), or averaged along each of the fitted
coronal loops (Figure 8). The ratios of the two quantities ranges from 
$\alpha_{\rm fit}/\alpha_{\rm loop}=0.99\pm0.00$
for the best case (\#7, Figure 8 top left) to $\alpha_{\rm fit}/\alpha_{\rm loop}=0.88\pm0.29$
for the worst case (\#12, Figure 6 bottom right). Our forward-fitting
code retrieves the correct sign of the $\alpha$-parameter in all cases,
and their absolute values agree within a few percents with the theoretical model.
Thus we conclude that the convergence behavior of our forward-fitting code
is quite satisfactory, because it retrieves the force-free $\alpha$-parameters
with high accuracy, at least for the given parameterization.

\section{	Forward-Fitting to Low and Lou (1990) Model  	}

The foregoing tests were necessary to verify how accurately the forward-fitting
code can retrieve the solution with many free parameters 
(from $n_{\rm free}=5,...,50$),
which represents a numerical convergence test. Of course, because the same
parameterization is used in simulating the input data as in the model that
is forward-fitted to the simulated data, this represents the most
favorable condition where the model parameterization is adequate for the input data.
Moreover, the simulated data were only force-free to second order, so we cannot
use the force-freeness figure of merit calculated from the solution as an absolute 
criterion to evaluate how accurate the forward-fitting solution fulfills Maxwell's
equations. So, the foregoing tests do not tell us whether the model 
parameterization of the forward-fitting code is adequate for arbitrary data,
and how physical the solution is. 

\begin{figure}
\centerline{\includegraphics[width=1.0\textwidth]{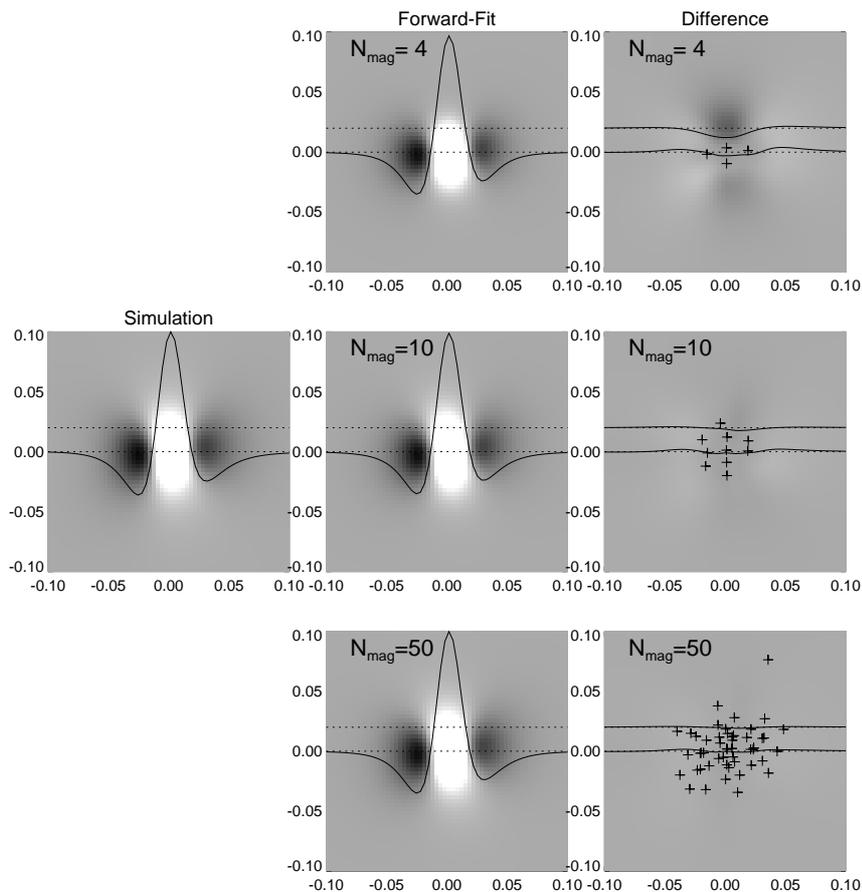}}
\caption{The decomposition of line-of-sight magnetogram $B_z(x,y)$ (simulation
in left middle frame) of the Low and Lou (1990) model is shown for three
trials with different numbers of magnetic components ($N_{\rm mag}=4, 10, 50$,
first, second, and thrid row). The locations of the center positions of the
magnetic components is shown with crosses in the difference images (right-hand 
panels). Two profiles across the middle of the magnetogram are also shown
(solid curves).}
\end{figure}

\begin{figure}
\centerline{\includegraphics[width=1.0\textwidth]{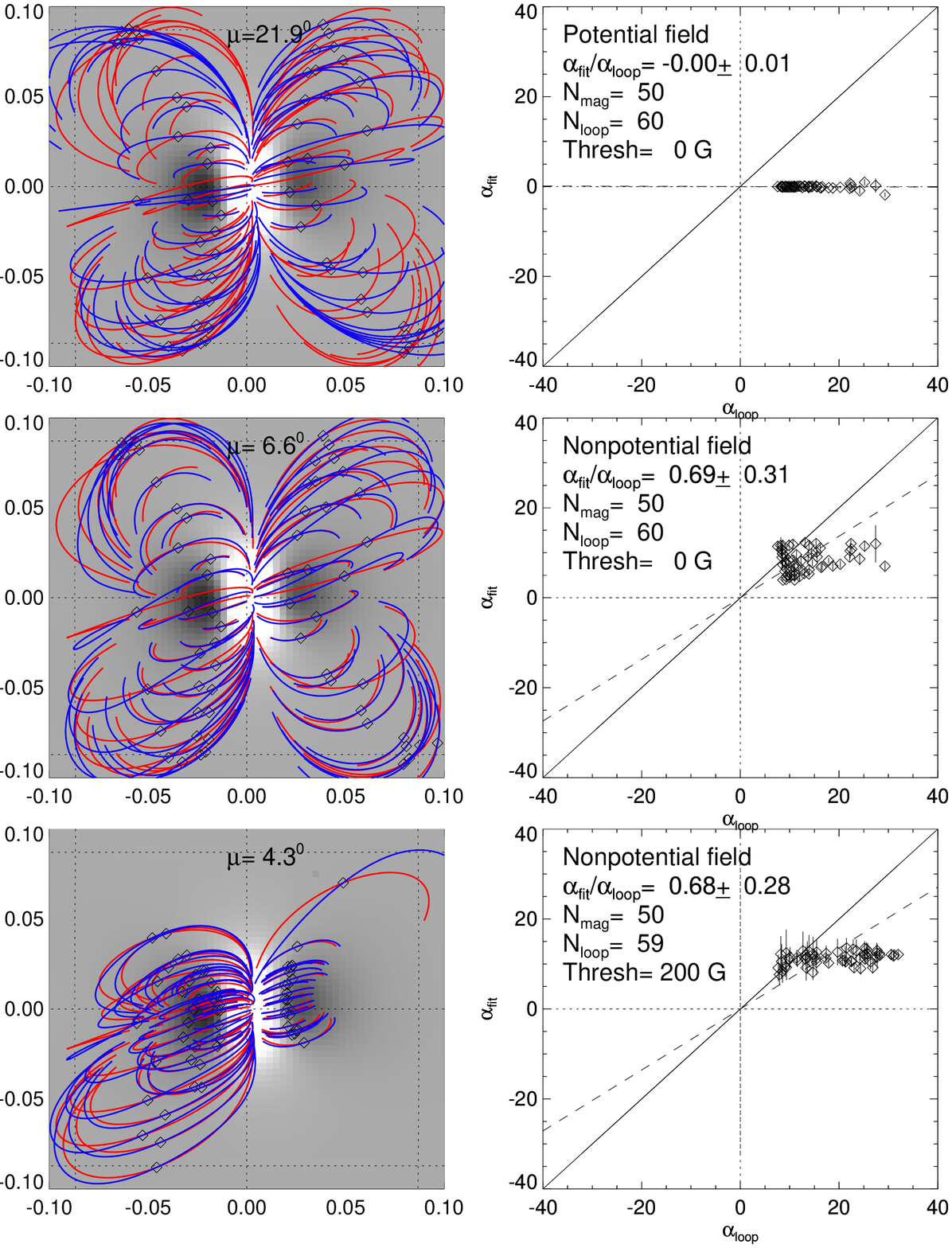}}
\caption{Potential field calculation (top) and forward-fitting of a
nonpotential (quasi-NLFFF) model (middle and bottom) to different sets 
(Thresh=0, 200 G) of $N_{\rm loop} \approx 60$ coronal loops, which represent 
an accurate nonlinear force-free field solution of the Low and Lou (1991) 
model. The model loops are outlined in red color, and the best-fit field
lines in blue color. The average misalignment angle $\mu$ is indicated in each
panel. The photospheric magnetogram is rendered with a greyscale.
A scatterplot of the best-fit $\alpha_{\rm fit}$-parameters averaged along each 
loop versus the model parameters $\alpha_{\rm model}$ are shown in the right-hand 
panels.}
\end{figure}

We conduct now a test that generates the input data with a completely 
different parameterization than our model and fit a non-potential field case 
that is exactly force-free, which is provided by analytical NLFFF solutions 
of the Low and Lou (1990) model, described and used also in Malanushenko 
\etal (2009). The particular solution we are using is defined by the 
parameters $(a=0.6, n=2.0)$, where $a$ is a Grad-Shafranov constant and $n$ 
is the harmonic number of the Legendre polynomial.

The line-of-sight magnetogram $B_z(x,y)$ of the Low and Lou case consists of
three smooth patches with an elliptical geometry, where the central patch has
a positive magnetic polarity, and the eastern and western patch a negative
polarity (see greyscale image in Figure 9 in left panel). The ideal number
of decomposed features in the magnetogram is not known a priori, because
a too small number leaves too large residuals of magnetic flux that is not
accounted for in the forward-fit, while a too large number leads to 
overlapping magnetic field components and force-free $\alpha$-parameter zones,
which may jeopardize the quality of forward-fitting (which works best for
spatially non-overlapping and independent zones). We show three different
trials with $N_{\rm mag}=4, 10, 50$ in Figure 9. The forward-fitted magnetograms
and the difference images with respect to the input magnetogram are also
shown in Figure 9. The residuals in the difference images have a mean and
standard deviation of $(B_{\rm fit}-B_{\rm model})/B_{\rm max}=
0.0022\pm0.0243$ for $N_{\rm mag}=4$; $-0.0005\pm0.0082$ for $N_{\rm mag}=10$;
and $-0.0016\pm0.0043$ for $N_{\rm mag}=50$, respectively. Thus, the
forward-fitted magnetograms agree with the Low and Low (1990) model
within $\lapprox 1\%$ of the magnetic flux. Note that the parameters that
decompose the line-of-sight magnetogram make up 80\% of the free parameters
in our forward-fitting model, fully determine the potential field 
extrapolation, but ignore the force-free $\alpha$-parameters so far.
The potential field solution for the Low and Lou (1990) model is shown
in Figure 10 (top panel), for a decomposition of $N_{\rm mag}=50$ magnetic
components, for a set of $N_{\rm loop}=60$ loops. The resulting mean
misalignment between the model and the potential field is $\mu=21.9^\circ$
(Table 6, first line), and $\mu=30.8^\circ$ for $N_{\rm mag}=10$, respectively.

\begin{table}
\caption{Best-fit results of forward-fitting to 
the Low and Lou (1990) model, using the following parameter settings:
the number of magnetic source compoonents $N_{\rm mag}=10, 50$, 
the threshold of the magnetic field for selected loops Thresh=0, 200 G, 
but otherwise standard settings as listed in Table 1.
The results are quantified by the number of fitted loops $n_{\rm loop}$,
the mean misalignment angle $\mu$ (degrees),
the ratio of the fitted to the model input force-free parameter,
$\alpha_{\rm fit}/\alpha_{\rm model}$, 
the divergence-freeness $L_{\rm d}$, the force-freeness $L_{\rm f}$,
and the computation time $t_{\rm CPU}$ (s).}
\footnotesize
\begin{tabular}{rrrrrrrr}
\hline
$N_{\rm mag}$ & Thresh & $n_{\rm loop}$ & $\mu$ & $\alpha_{\rm fit}/\alpha_{\rm model}$ & 
						$L_{\rm d}$ & $L_{\rm f}$ &$t_{\rm CPU}$\\
          & [G]      &            & (deg) &          &       &       & (s)    \\
\hline
50 &   0 & 60  & 21.9$^\circ$ &  $0.00\pm 0.02$ & 0.000021 & 0.000023 &    0 \\
10 &   0 & 60  & 12.7$^\circ$ &  $0.66\pm 0.43$ & 0.000083 & 0.000751 &  257 \\
50 &   0 & 60  &  6.6$^\circ$ &  $0.69\pm 0.31$ & 0.000045 & 0.000082 & 1359 \\
10 & 200 & 59  &  6.1$^\circ$ &  $0.63\pm 0.22$ & 0.000121 & 0.000617 &  123 \\
50 & 200 & 59  &  4.3$^\circ$ &  $0.68\pm 0.28$ & 0.000084 & 0.000174 & 1338 \\
\hline
\end{tabular}
\end{table}

We forward-fitted several hundred runs to the Low and Lou (1990) model
with different parameter settings (Table 1) and list the results of a
selection of four cases in Table 6, and two cases thereof in Figure 10.
For $N_{\rm mag}=50$ and a threshold of Thresh=0 G we find a solution
that has only a misalignment of $\mu=6.6^\circ$ (Figure 10, middle panel,
and Table 6, third line). This case retrieves the force-free parameter
$\alpha$ with an average ratio of $\alpha_{\rm fit}/\alpha_{\rm model}
=0.69\pm0.31$ (Figure 10, middle right panel) for the 60 loops shown.
The divergence-freeness and force-freeness amount to $L_{\rm d}=4.5 \times
10^{-5}$ and $L_{\rm f}=8.2 \times 10^{-5}$. If we select a set of coronal
loops with only strong magnetic field strengths at the footpoints 
(Thres=200 G),
the misalignment improves to $\mu=4.3^\circ$ (Figure 10, bottom left panel), 
while the accuracy of the retrieved $\alpha$-values remains about the same 
($\alpha_{\rm fit}/\alpha_{\rm model}=0.68\pm0.28$ (Figure 10, bottom right panel).
It appears that our forward-fitting code always underestimates the
values in loops with the highest $\alpha$-parameter, which was not the
case in all of our previous simulations (Figure 8). It appears that the
elliptical shape of magnetic patches could be responsible for this
underestimate, while it did not occur for spherical shapes of magnetic
patches (Simulation runs \#7-12) described in Section 4. Nevertheless,
the achieved small amount of misalignment down to $\mu=4.3^\circ$ 
yields a good approximation to a nonlinear force-free field that is
sufficiently accurate for most practical purposes of coronal field
modeling and can be obtained in a relatively short computation time.
The computation times for the five runs listed in Table 6 amounted to
$t_{\rm CPU} \approx 2-20$ minutes. We obtained even higher accuracies down
to misalingmens of $\mu \lapprox 1^\circ$ for smaller subgroups of
coronal loops that were localized in partial domains of the active region.

\section{	Discussion and Conclusions  		  }

In this study we developed a numeric code that accomplishes
(second-order) nonlinear force-free field fast forward-fitting
of combined photospheric magnetogram and coronal loop data.
The goal of this code is to compute a realistic magnetic field of a
solar active region. Previously developed magnetic field extrapolation 
codes used either photospheric data only, such as {potential-field
source surface (PFSS)} codes (\eg Altschuler and Newkirk, 1969) and 
{\sl nonlinear force-free field (NLFFF)} codes (\eg 
Yang \etal, 1986; Wheatland \etal, 2000, 2006; Wiegelmann, 2004; 
Schrijver \etal, 2005, 2006; Amari \etal, 2006; Valori \etal, 2007; 
Metcalf \etal, 2008; DeRosa \etal, 2009; Malanushenko \etal, 2009),
or (stereoscopically triangulated) coronal loop data only
(Sandman \etal, 2009; Sandman and Aschwanden, 2011). There are only
very few attempts where both photospheric and coronal data constraints
were used together to obtain a magnetic field solution, using either
a potential field model with unipolar buried charges that could be
forward-fitted to the observed loops (Aschwanden and Sandman, 2011), 
a linear force-free field (Feng \etal, 2007a,b), or a NLFFF code
(Malanushenko \etal, 2009, 2011). For special
geometries, potential field stretching methods (Gary and Alexander, 1999)
or a minimum dissipative rate method for non-forcefree fields 
have also been explored (Gary, 2009). 

The new approach
of including coronal magnetic field data, in form of stereoscopically
triangulated loop 3D coordinates, requires a true forward-fitting approach, 
while the traditional use of photospheric magnetogram (or vector magnetograph)
data represents an extrapolation method from given boundary constraints.
Both methods require numerous iterations, and thus are computing-intensive,
but the classical forward-fitting method requires a suitable parameterization
of a magnetic field model, while extrapolation methods put no constraints
on the functional form of the solutions (such as the 3D geometry of
magnetic field lines). 
Thus, the new approach developed here makes use of a parameterization 
of the 3D magnetic field model in terms of analytical functions that can be
fitted relatively fast to the given coronal constraints, but may lack
the absolute generality of nonlinear force-free field solutions that
NLFFF codes are providing. However, our analytical NLFFF model, which is 
accurate
to second-order (Paper I), probably represents one of the most general 
parameterizations that is possible with a minimum of free parameters,
adapted to uniformly twisted field lines. The parameter space
given by this model represents a particular class of quasi-forcefree
solutions, which is supposed to be most suitable for a superposition of
twisted field line structures, but only fitting to real data can reveal
how useful and suitable our model is for applications to solar data.

In this study we described the numeric code, which is based on the
analytical second-order solutions derived in Paper I, and performed
test with 12 simulated cases (six potential and six non-potential), 
as well as with an analytical NLFFF solution of the Low and Lou (1990)
model. The forward-fitting to the 12 simulated cases demonstrated
(i) the satisfactory convergence behavior of the forward-fitting code
(with mean misalignment angles of $\mu=3.4^\circ\pm2.1^\circ$ for
potential field cases (see Table 2), and $\mu=5.1^\circ\pm4.3^\circ$ 
for non-potential field cases (see Table 4), (ii) the relatively
fast computation speed (from $\lapprox 1$ s to $\lapprox 10$ min);
and (iii) the high fidelity of retrieved force-free $\alpha$-parameters
($\alpha_{\rm fit}/\alpha_{\rm model} \approx 0.9-1.0$; see Figure 8).
The additional test of forward-fitting to the analytical solution of
Low and Lou (1990) data yielded similar results, \ie
satisfactory convergence behavior 
(with mean misalignment angles of $\mu=4.3^\circ-6.6^\circ$ for
two subsets of loops, see Figure 10), (ii) relatively fast computation speed 
($t_{\rm CPU} \approx 2-20$ min); and (iii) the fidelity of retrieved 
force-free $\alpha$-parameters ($\alpha_{\rm fit}/\alpha_{\rm model} 
\approx 0.7\pm0.3$; see Figure 10). The only significant difference 
of the second test is the trend of underestimating the
$\alpha$-parameter for those loops with the highest $\alpha$-values,
by a factor of $\gapprox 0.5$. 
However, if the loops with the highest
$\alpha$-values are fitted individually, the code retrieves the
correct $\alpha$-value. It is not clear whether this feature of the code
is related to the geomerical shape of the magnetic concentrations in
the magnetogram, which is spherical in our simulation and forward-fitting
model, but elliptical in the Low and Lou (1990) case. We simulated the
elliptical magnetic sources of the Low and Lou (1990) model by a
superposition of spherical sources and found that the code retrieves
the correct $\alpha$-values for each loop (within a few percent accuracy).
It is possible that the geometric shape of the Low and Lou (1990) model,
which represents a special class of nonlinear force-free solutions
anyway (in terms of Legendre polynomials) cannot efficiently be
parameterized with a small number of spherical components, which is
the intrinsic parameterization of our code. 
Anyway, since the Low and Lou (1990) model represents
also a very special subclass of nonlinear force-free solutions that 
may or may not be adequate to model real solar active regions, 
it may not matter much for the performance of our code with real solar data.

After having tested our numeric code we can proceed to apply it to
solar data, such as active regions observed with STEREO since 2007, for
which stereoscopic triangulation of coronal loops is available
(Feng \etal, 2007a,b; Aschwanden and Sandman, 2010; Aschwanden \etal, 2012a,b). 
The second-order NLFFF approximations of our code may 
be used as an initial guess for other more accurate NLFFF codes,
resulting into a significantly shorter computation time. Other future
developments may involve the reduction of coronal constraints from
3D to 2D coordiantes, which can be furnished by automatic loop
tracing codes (\eg Aschwanden \etal, 2008; Aschwanden, 2010; and 
references therein) and does not require the availability of STEREO data. 
However, non-STEREO data provide less rigorous constraints for coronal loop
modeling, and thus increase the ambiguity of force-free field solutions.
Nevertheless, more realistic coronal magnetic field models seem now in
the grasp of our computation methods, which has countless benefits for
many research problems in solar physics.

\acknowledgements
Part of the work was supported by
NASA contract NNG 04EA00C of the SDO/AIA instrument and
the NASA STEREO mission under NRL contract N00173-02-C-2035.

\section*{References} %%% REFERENCES 

\def\ref#1{\par\noindent\hangindent1cm {#1}}

\small
\ref{Altschuler, M.D., Newkirk, G.Jr.: 1969, \sp {\bf 9}, 131.}
\ref{Amari, T., Boulmezaoud, T.Z., Aly, J.J.: 2006, \aap {\bf 446}, 691.} 
\ref{Aschwanden, M.J.: 2004, {\sl Physics of the Solar Corona. An Introduction},
        Praxis Publishing Co., Chichester UK, and Springer, Berlin, 
	Section 5.3.}
\ref{Aschwanden, M.J., Lee, J.K., Gary, G.A., Smith, M., Inhester, B.:
 	2008, \sp {\bf 248}, 359.}
\ref{Aschwanden, M.J.: 2010, \sp {\bf 262}, 399.}
\ref{Aschwanden, M.J., Sandman, A.W.: 2010, \aj {\bf 140}, 723.}
\ref{Aschwanden, M.J.: 2012, \sp this volume; Paper I.}
\ref{Aschwanden, M.J., W\"ulser, J.P., Nitta, N.V., Lemen, J.R., DeRosa, M., 
	Malanu\-shenko, A.: 2012a, \apj submitted.}
\ref{Aschwanden, M.J., W\"ulser, J.P., Nitta, N.V., Lemen, J.R.: 2012b, \sp in press.}
\ref{DeRosa, M.L., Schrijver, C.J., Barnes, G., Leka, K.D., Lites, B.W.,
        Aschwanden, M.J., \etal : 2009, \apj {\bf 696}, 1780.}
\ref{Feng, L., Inhester, B., Solanki, S., Wiegelmann, T., Podlipnik, B.,
        Howard, R.A., W\"ulser, J.P.: 2007a, \apjl {\bf 671}, L205.}
\ref{Feng, L., Wiegelmann, T., Inhester, B., Solanki, S., Gan, W.Q., 
	Ruan, P.: 2007b, \sp  {\bf 241}, 235.}
\ref{Gary, A., Alexander, D.: 1999, \sp {\bf 186}, 123.}
\ref{Gary, G.A.: 2009, \sp {\bf 257}, 271.}
\ref{Grad, H., Rubin, H.: 1958, {\it Proc. 2nd UN Int.~Conf.~Peaceful Uses
	of Atomic Energy}, {\bf 31}, 190.}
\ref{Low, B.C., Lou, Y.Q.: 1990, \apj {\bf 352}, 343.}
\ref{Malanushenko, A., Longcope, D.W., McKenzie, D.E.: 2009,
        \apj {\bf 707}, 1044.}
\ref{Malanushenko, A., Yusuf, M.H., Longcope, D.W.: 2011,
        \apj {\bf 736}, 97.}
\ref{Metcalf, T.R., Jiao, L., Uitenbroek, H., McClymont, A.N., 
	Canfield,R.C.: 1995, \apj {\bf 439}, 474.}
\ref{Metcalf, T.R., DeRosa, M.L., Schrijver, C.J., Barnes, G., 
	van Ballegooijen, A.A., Wiegelmann, T., Wheatland, M.S.,
	Valori, G., McTiernan, J.M.: 2008, \sp {\bf 247}, 269.}
\ref{Press, W.H., Flannery, B.P., Teukolsky, S.A., Vetterling, W.T.:
 	1986, {\sl Numerical Reci\-pes, The Art of Scientific Computing},
 	Cambridge University Press, Cambridge, 294.}
\ref{Ruan, P., Wiegelmann, T., Inhester, B., Neukirch, T., Solanki, S.K.,
        Feng, L.: 2008, \aap {\bf 481}, 827.}
\ref{Sandman, A., Aschwanden, M.J., DeRosa, M., W\"ulser, J.P., 
	Alexander, D.: 2009, \sp {\bf 259}, 1.}
\ref{Sandman, A.W., Aschwanden, M.J.: 2011, \sp \bf{270}, 503.} 
\ref{Schrijver, C.J., DeRosa, M.L., Metcalf, T.R., Liu, Y., McTiernan, J.,
	Regnier, S., Valori, G., Wheatland, M.S., Wiegelmann, T.:
 	2006, \sp {\bf 235}, 161.}
\ref{Schrijver, C.J., DeRosa, M.L., Metcalf, T.,  Barnes, G., Lites, B., 
	Tarbell, T., \etal : 2008, \apj {\bf 675}, 1637.}
\ref{Valori, G., Kliem, B., Fuhrmann, M.: 2007, \sp {\bf 245}, 263.}
\ref{Wheatland, M.S., Sturrock, P.A., Roumeliotis, G.:
 	2000, \apj {\bf 540}, 1150.}
\ref{Wheatland, M.S.: 2006, \sp {\bf 238}, 29.}
\ref{Wiegelmann, T., Neukirch, T.: 2002, \sp {\bf 208}, 233.}
\ref{Wiegelmann, T., Inhester, B.: 2003, \sp {\bf 214}, 287.}
\ref{Wiegelmann, T.: 2004, \sp {\bf 219}, 87.}
\ref{Wiegelmann, T., Lagg, A., Solanki, S.K., Inhester, B., Woch, J.:
        2005 \aap {\bf 433}, 701.}
\ref{Wiegelmann, T., Inhester, B.: 2006, \sp {\bf 236}, 25.}
\ref{Yang, W.H., Sturrock, P.A., Antiochos, S.K.: 1986, \apj {\bf 309}, 383.}

\end{article}
\end{document}